\documentclass[useAMS,usenatbib]{mn2e}
\usepackage[dvips]{graphicx}
\usepackage{longtable}
\usepackage{epsfig}
\usepackage[colorlinks,urlcolor=blue]{hyperref}
\usepackage{amssymb,amsfonts,amsmath,mathtext,cite,enumerate,float}
\usepackage{xcolor}
\usepackage{multirow}
\usepackage{textcomp}

\title[GW150914 --- II. MASTER  Optical Follow-Up observations. ]{\large{First Gravitational-Wave Burst  GW150914: Part II. MASTER Optical Follow-Up Observations}}
\author[V. M. Lipunov et al.]{\small{V. M. Lipunov$^{1,2}$\thanks{E-mail: lipunov2007@gmail.com (VML)}, V.Kornilov$^{1,2}$, E.Gorbovskoy$^{2}$, N.Tiurina$^{2}$, P.Balanutsa$^{2}$, A.Kuznetsov$^{2}$},
\newauthor \small{V.Vladimirov$^{2}$, D.Vlasenko$^{1,2}$, I.Gorbunov$^{1,2}$, V.Chazov$^{2}$, D.Kuvshinov$^{1}$, A.Gabovich $^{2}$},
 \newauthor \small{D.A.H. Buckley,$^{3}$, S.B.Potter $^{3}$, A.Kniazev$^{3}$, S. Crawford$^{3}$, R. Rebolo Lopez$^{4}$, M. Serra Ricart$^{4}$, G. Israelian $^{4}$, N. Lodieu $^{4}$},
   \newauthor \small{O. A. Gress$^{5}$, N.M.Budnev$^{5}$, K. I. Ivanov$^{5}$, V.Poleschuk$^{5}$, S.Yazev$^{5}$, A.Tlatov$^{6}$, V.Senik$^{6}$, D.Dormidontov$^{6}$},
       \newauthor \small{A.Parkhomenko$^{6}$, V. Yurkov $^{7}$, Yu. Sergienko $^{7}$,},
 \newauthor \small{
   R. Podesta$^{8}$,
   H. Levato$^{9}$,
   C. Lopez$^{8}$,
C. Saffe$^{9}$,
C. Mallamaci$^{8}$}
\\
\\
$^{1}$M.V.Lomonosov Moscow State University, Physics Department, Leninskie gory, GSP-1, Moscow, 119991, Russia \\
$^{2}$M.V.Lomonosov Moscow State University, Sternberg Astronomical Institute, Universitetsky pr., 13, Moscow, 119234, Russia\\
$^{3}$ South African Astronomical Observatory, PO Box 9, 7935 Observatory, Cape Town, South Africa\\
$^{4}$Instituto de Astrofacuteisica de Canarias Via Lactea, s/n E38205 - La Laguna (Tenerife), Spain\\
$^{5}$Irkutsk State University, Applied Physics Institute, 20, Gagarin blvd,664003, Irkutsk, Russia\\
$^{6}$Kislovodsk Solar Station of the Main (Pulkovo) Observatory RAS, P.O.Box 45, ul. Gagarina 100, Kislovodsk 357700, Russia\\
$^{7}$Blagoveschensk State Pedagogical University, Lenin str., 104, Amur Region, Blagoveschensk 675000, Russia\\
$^{8}$Observatorio Astronmico F´elix Aguilar, Av. Benavidez 8175(O), 5413 San Juan, Argentina\\
$^{9}$Instituto de Ciencias Astronomicas, de la Tierra y del Espacio, Av.España Sur 1512, J5402DSP, San Juan, Argentina}
\begin{document}

\date{Accepted 2016 XXX. Received 2016 ; in original form 2016 March}
\pagerange{\pageref{firstpage}--\pageref{lastpage}} \pubyear{2016}
\maketitle
\label{firstpage}

\begin{abstract}
\footnotesize{The Advanced LIGO observatory recently reported the first direct detection of gravitational waves predicted by Einstein (1916). We report on the first optical observations of the Gravitational Wave (GW) source GW150914 error region with the Global MASTER Robotic Net.  We detected several optical transients, which proved to be unconnected with the GW event. Our result is consistent with the assumption that gravitational waves were produced by a binary black hole merger. The detection of the event confirmed the main prediction of the population synthesis performed with the ``Scenario Machine'' formulated in Lipunov1997b.
}
\end{abstract}



\section{Introduction}

The Advanced LIGO observatory recently reported the first direct detection of gravitational waves \citep{Abbott2016a,Abbott2016c} as a merger of two black holes with the mass of $36^{+5}_{−4} M_{\bigodot}$ and $29^{-4}_{−4} M_{\bigodot}$.

There are several arguments that electromagnetric (EM) radiation should appears before, during and after Gravitational Wave (GW) event.   ~\citet{Lipunov1996a} showed that if the merging process involves at least one magnetized neutron star, one can expect short radio and optical precursor nonthermal emission, like that produced by pulsars. ~\citet{Hansen} later illustrated the idea by ~\citet{Lipunov1996a}  for the case of a detailed electrodynamic model. \citet{Blinnikov} were the first to show that a neutron star merging can be accompanied by a powerful electromagnetic burst. After the merging a part of the radioactive matter can be ejected leaving behind the so-called Kilonova ~\citep{Li,Metzger,Tanvir,Berger} or a rapidly rotating self-gravitating object - magnetorotational Spinar  ~\citep{Lipunova,Lipunov2008} - may form.
We also do not rule out the possibility of a gamma-ray burst whose electromagnetic radiation is concentrated in a narrow jet ~\citep{Eichler,Narayan}, which is very unlikely to be detected during the GW event due to the low probability that it is beamed towards the Earth.

A  Scenario Machine Prediction of the binary black holes merging as the first events that would be discovered at gravitational interferometers \citep{Lipunov1997a,Lipunov1997b,Lipunov1997c} and  theoretical analysis  are given in the first part of this work \citep{MASTER_LIGO_Theory}. Here we will focus in detail on the optical  follow-up observation of the first in the human history gravitational-wave event GW150914 by the MASTER Global Robotic Net.

Starting from the 2003 we began to develop a program of robotized observations of gamma-ray bursts and another burst-like phenomena (optical transients), see MASTER project description, \citet{Lipunov2005,Lipunov2010,Lipunov2016}, with the primary aim to perform optical observations of Gamma Ray Bursts (GRBs).  We developed the MASTER Global network of identical twin-tube colored wide-field telescopes with real-time reduction deployed both in the Northern and Southern hemispheres ~\citep{Lipunov2010,Kornilov2012,Gorbovskoy2012}.

This led us to join the LVC follow-up program in 2015 to detect possibly optical counterparts of GW events ~\citep{Abbott2016b}.

On 16 September 2015 at 19h 47s UT we obtained the probability matrix for the error box of the first gravitational-wave alert ALIGO trigger, G184098 ~\citep{GCN18330}. Starting from the following night we began inspecting the probable GW event sky areas with  MASTER network telescopes at the following sites: MASTER-Amur, MASTER-Tunka, MASTER-Kislovodsk, MASTER-SAAO and MASTER-IAC, where the weather and night-time conditions were permitted. We monitored about 5000 square degrees of sky with different depths, down to a limiting magnitudes as faint as 20m. These results are partically reviewed in a paper by the LIGO/VIRGO EM collaboration \citet{Abbott2016b}.

\section{MASTER Global Robotic  Net }
\begin{figure*}
\centering
\begin{minipage}{170mm}
\center{\includegraphics[width=1\linewidth]{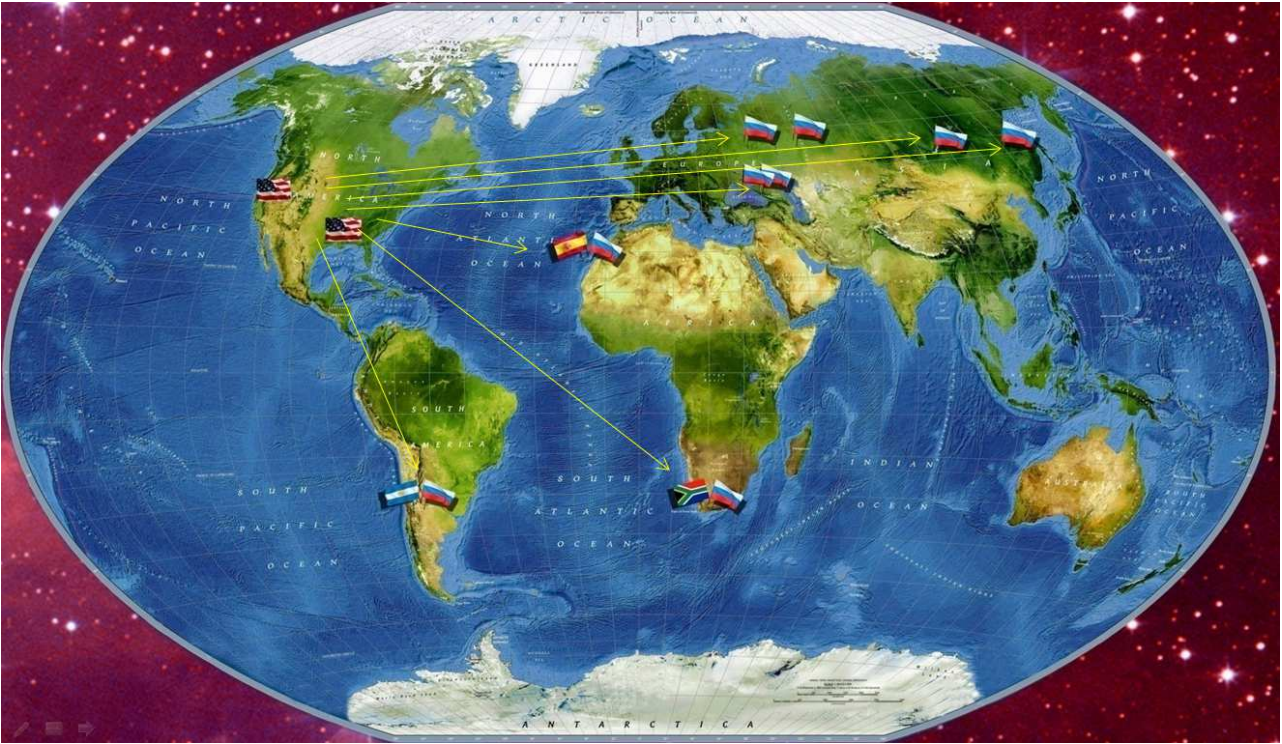}}
\caption{Global MASTER robotic telescope sites and LIGO interferometers, that were involved into GW150914 event's investigations. Information about each MASTER-NET observatory available at Table \ref{tab:masternet}.  }
\end{minipage}\label{Position}
\label{fig_MASTER_NET}
\end{figure*}

The MASTER Global Robotic Net\footnote{\url{http://observ.pereplet.ru/}} includes several observatories with identical instruments: MASTER-Amur,  MASTER-Tunka, MASTER-Ural, MASTER-Kislovodsk (Russian Federation), MASTER-SAAO (South Africa) and MASTER-IAC (Spain, Canarias), and Very Wide Field cameras (MASTER-VWF) in Argentina~\citep{Lipunov2004,Lipunov2010,Kornilov2012,Gorbovskoy2013}, see  Figure \ref{fig_MASTER_NET} and Table \ref{tab:masternet}. Each MASTER observatory  provides a survey speed of 128~deg$^2$ per hour with a limiting magnitude (in white light) of 19 or 20, respectively,  on grey or dark nights. Each observatory is equipped with two wide field 40cm optical telescopes (MASTER-II) and two very wide field optical cameras (MASTER-VWF). Each MASTER II system consists of twin telescopes, each with a total field of view of 4~deg$^2$ and with a 4098~pixel~$\times$~4098~pixel CCD camera with a scale of 1.85$''$/pix, and with \textit{BVRI} and polarizing filters. It is possible to observe without any filter in integral (white) light with or without two orthogonal polarizers. Details of the MASTER filters and polarization measurements can be found in \citet{Kornilov2012} and \citet{Gorbovskoy2013}.

The observations with MASTER-Net can be performed in different modes: alert, survey, and inspection. The alert mode is initiated if a target position has good accuracy (when the error-box size is less than  4$^\circ$, which is the field of view of each of the MASTER twin telescopes, and is usually used to observe GRBs upon receiving notices from the Gamma-ray Coordinates Network\footnote{\url{http://gcn.gsfc.nasa.gov/}} (GCN), neutrino alerts or gravitational wave’s alerts.

MASTER observes GRB alerts with error-boxes less then 2 deg$^2$ in the alert mode, with co-aligned tubes and different polarizers (total of 4~deg$^2$). In the case of alerts with larger error boxes (e.g. Fermi gamma ray alerts; GW alerts, etc.), MASTER observations are performed with the twin telescopes off-set to cover $2^{\circ} \times 4^{\circ}$ , i.e. 8 deg$^2$, imaging 3 exposures per field \citep{Lipunov2016,Gorbovskoy2016}.

MASTER survey mode is used for the regular survey and search for optical transients (OTs), when there are no alerts, and is the usual mode of operation. MASTER control and planning software have been developed to select  preferred locations for the survey. The planner takes into account the previous coverage rate of the area, angular distances from the Galactic plane, the Moon, the Sun, the ecliptic, and from the current \textit{Swift} FOV. It takes into account the number of SNe~Ia in the field and GRBs discovered within the previous 24 hours. During the survey each area is observed several times with automatically chosen exposure times ranging from 60 to 180~s, depending on the presence of the Moon. The observation time varies from 10~min to 1~h, depending on the Moon phase, weather conditions, and the remaining observing time.
In the case of large coordinate error-boxes, which is the case more than 1h after the trigger, we use the inspection mode, which combines the alert and survey modes. First, the centre of the error box is observed in the alert mode during the time  $t-T_{0} < 5$ min. Then the telescope switches to the survey mode inside the error box area. The 1-$\sigma$ error box is covered first, then larger 2-$\sigma$ and 3-$\sigma$ regions. The error boxes are covered using the same algorithm as for the survey mode. Each area is observed three times with five minute intervals with exposure times 60~s. The inspection mode allows us to cover big areas quickly and search for all types of OTs.  If the same error-box can be observed by two or more MASTER Net telescopes, then they are commanded to cover different fields. Thus, the rate of coverage grows in proportion to the number of telescopes.
The main unique feature of MASTER system is our dedicated software developed over 10 years, which allows new optical transients to be discovered in MASTER images within 1-2 minutes after each CCD exposure.
This information includes the full classification of all sources found in the image, the data from previous MASTER-Net archive images for each source, full information from the VIZIER database and all public sources (e.g. Minor Planet checker), derivation of orbital elements for moving objects, etc.
For transient detections, real astrophysical sources are unlikely to be represented by just 1, 2 or 4 pixels in the image, such sources are very likely to be artificial and are screened out by the transient search task.  Real transient sources must have more than 10 pixels distributed in accordance with a specific profile to distinguish them from hot pixels. MASTER software discovers optical transients not by the differencing between the previous and current frames, but by fully identifying each new source in every frame, with respect to a reference image.
If there is a galaxy in the neighborhood of a transient, the software automatically checks for this and classifies the OT as a possible SNe (PSN), after manually checking its position to find any faint Galactic source that is below the optical frame limit along the line of sight in MASTER or POSS archive images.  An inspection of MASTER's Andromeda image\footnote{\url{http://observ.pereplet.ru/MASTER-M31.jpg}} reveals possible novae in M31.  If there are no VIZIER sources within 5'' and the brightness is constant over 1-2 nights, it can be a possible cataclysmic variable (mostly of the dwarf nova type). If the brightness increases and fades away again over the course of several tens of minutes and there is a red or infrared detection in VIZIER, it is likely to be a dMe flare star (UV Cet) object.
The discovery strategy for optical transients consists in the following. The objects detected in a MASTER image can be classified into three categories:
\begin{enumerate}

\item  Known objects - these objects are identified by matching their coordinates and magnitude with catalogues;
\item  $"$flare$"$ -  the object is found at the location as a catalogued object, but has a significant magnitude difference (either negative or positive);
\item  Unknown - the object is absent in the catalogues.

\end{enumerate}
We then compare the object lists to filter out uncatalogued moving objects and to start analyzing the transients found.
This MASTER software has allowed us to discover more than $\sim$1000 optical transients of ten different types (GRB optical counterparts, supernovae (including the superluminous ones), novae, QSO and blazar flares, short transients (possible orphan GRBs), dwarf novae, antinovae($\varepsilon$ Aur-type \citep{Lipunov2016AA}, R CrB and other cataclysmic variables (VY Scl-type), UV Cet type f;are stars, potentially hazardous asteroids and comets) \footnote{\url{http://observ.pereplet.ru/MASTER\_OT.html}} for several years.
This is a fully automatic detection system that takes 1-2 minutes after the CCD readout to analyze the frame. After automated OT detection and primary classification, each candidate is carefully analyzed by a human (typically within 24 h) to further investigate its nature. If we have several images of the OT, we analyze its light curve (LC) and MASTER archive images. Then a human also analyses the public databases (like VIZIER) in this area. If the error box of the alert has been imaged, and there are no sources in previous or archived MASTER images (lists of objects), and no sources found within 5" in the VIZIER database, then the OT is a likely new discovery.

\section{GW~150914 observations}

The GW150914 alert message was received more than one day after the GW event. All telescopes of the MASTER network began observing different parts of the GW150914 error region when the corresponding areas became visible. The first images in response to the GW150914 alert were taken at MASTER-SAAO observatory at 2015-09-16 20:18:11 UT.
The initial Ligo error region consisted of two elongated areas. The first area was located in the Southern Hemisphere and the second near the celestial equator. Both areas were somewhat difficult to observe. The two areas rose only several hours before the sunrise. In addition, most of the error region was less than 40$^{\circ}$ from the Sun, where no regular survey-mode observations are performed with MASTER telescopes.

The southern GW150914 localization area was observed with the MASTER-SAAO telescope located in the Southern Hemisphere (SAAO; South Africa). The area near the equator was observed by MASTER-IAC, MASTER-Kislovodsk, MASTER-Tunka, and MASTER-Amur telescopes located in the Northern Hemisphere.

\begin{figure*}
\centering
\center{\includegraphics[width=1\linewidth]{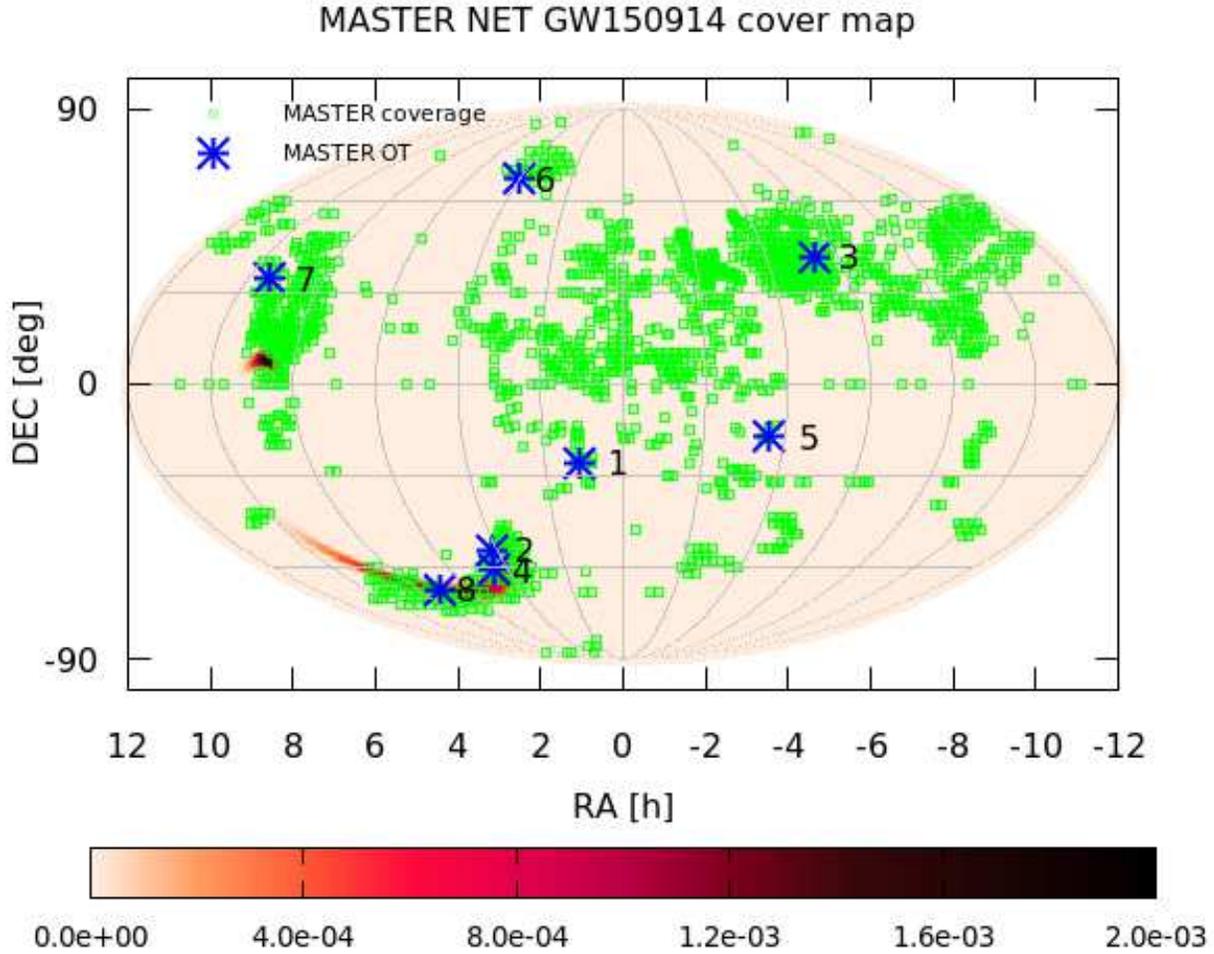}}
\caption{The complete map of the sky survey carried out by MASTER robotic telescope net at the time of the GW150914 observation, from 14 to 22 September 2015. Each field (marked with a green square) was observed at least 3 times and covers 4 square degrees  of the sky down to a limiting magnitude of 19-20$^m$. The color palette indicates Ligo GW150914 probability distribution over the sky. The probability is nowhere zero, and therefore any field can be considered. The blue asterisks show the optical transients discovered by MASTER during the inspections of the Ligo error box and are described in the paper and details presented in Table \ref{tab:ligo1trans} }
\label{fig_Ligo_Cover1}

\end{figure*}

The MASTER-SAAO  twin robotic telescope of the Global MASTER Robotic Net  ~\citep{Lipunov2010} started inspecting the aLIGO trigger G184098 error box  61.25 h after the trigger time, at 2015-09-16 20:18:11 UT, after receiving the notice letter at 05:39:58 on September 16, 2015, later published in GCN18330.
We later checked MASTER’s database for earlier images taken on September 14, 15, and 16, 2015. We have 30 images taken, starting from 2015-09-15 03:24:22 UT, during the usual MASTER-SAAO survey. These images cover 16 square degrees (the stacked limit is 19.0m). So the first optical images were obtained by MASTER 1.094d before the notice letter and 17.6 h after the G184098 trigger.
During the inspection of GW150914  (so called aLIGO trigger G184098), the 5-sigma upper limit on our  sets was about 18.4 mag - 19.9 mag ~\citep{GCN18333,GCN18390}. On this first night we observed 212 deg$^2$, imaged 3 times for each field during $\sim$2 hours. The LMC and Milky Way are near the center and east edge of the error region, respectively. The coverage map is presented in Figure \ref{fig_Ligo_Cover1} and Figure \ref{MASTER_FERMI} and will be dicussed later.  We started surveying from the left edge of the error region as it rose.

MASTER-SAAO and other telescopes of MASTER net continued to survey the error region over the coming days. Up to the 22 September 2015, we took about 9500 images which is covered more than 5200 deg$^2$ of sky.  More than 920 images were located inside the eventual error region of GW150914 and cover 590 deg$^2$. Each area was covered several times. The full coverage map is shown in Figure \ref{fig_Ligo_Cover1}. The total probability of source location in the covered fields depends on the specific error region and reaches 56\%.  The values for all error boxes and for all observatories of MASTER net listed in Table \ref{tab:MASTER_Prob}

As follows from the table \ref{tab:MASTER_Prob} the most coverage of \textbf{any} GW150914 error region was by MASTER-SAAO. We observed locations of the GW150914 (G184098) since 2015-09-15 03:24:22UT  (in regular sky survey, before the GW trigger) until  2015-09-22 03:25:02 on every night for $\sim$2 hours before sunrise, when the field reached a 15$^{\circ}$ altitude at this site \citep{GCN18903}. We missed the fullnight of 2015-09-18 due to bad weather. Our observations covered about a half of the full probability area of the final error region, down to a limiting magnitude of ~19.0.

\section{Optical transients}

Our survey revealed eight optical transients observed at different observatories of MASTER network during this period. We mark all the newly discovered optical transients by the blue asterisks in Figure 2  and list them in Table \ref{tab:ligo1trans} with brief comments. Five of these eight optical transients are located in areas with very low probability (which, however, is greater than zero in all cases). The probability of their association with the gravitational wave source is extremely low.

The remaining three optical transients are located inside the error region. These OTs are marked by bold font and asterisks in Table \ref{tab:ligo1trans} and Figure \ref{fig_Ligo_Cover1}, respectively.  We discuss these in more detail below.

\subsection{MASTER OT J070747.72-68s -541316.9: A Possible Supernova  }
\label{sec:sn}
The MASTER-SAAO auto-detection system discovered an OT source at (RA, Dec) = 04h 09m 38.68s -54d 13m 16.9s on 2015-09-16.87912 UT ~\citep{Gress2015a}. The OT unfiltered magnitude was 17.3m (the limit is 17.8m). The OT was seen in 5 images. There are no minor planets at this location. We have a reference image without the OT taken on 2015-02-14.89772 UT with an unfiltered limiting magnitude of 20m.

\begin{figure*}
\centering
\center{\includegraphics[width=1\linewidth]{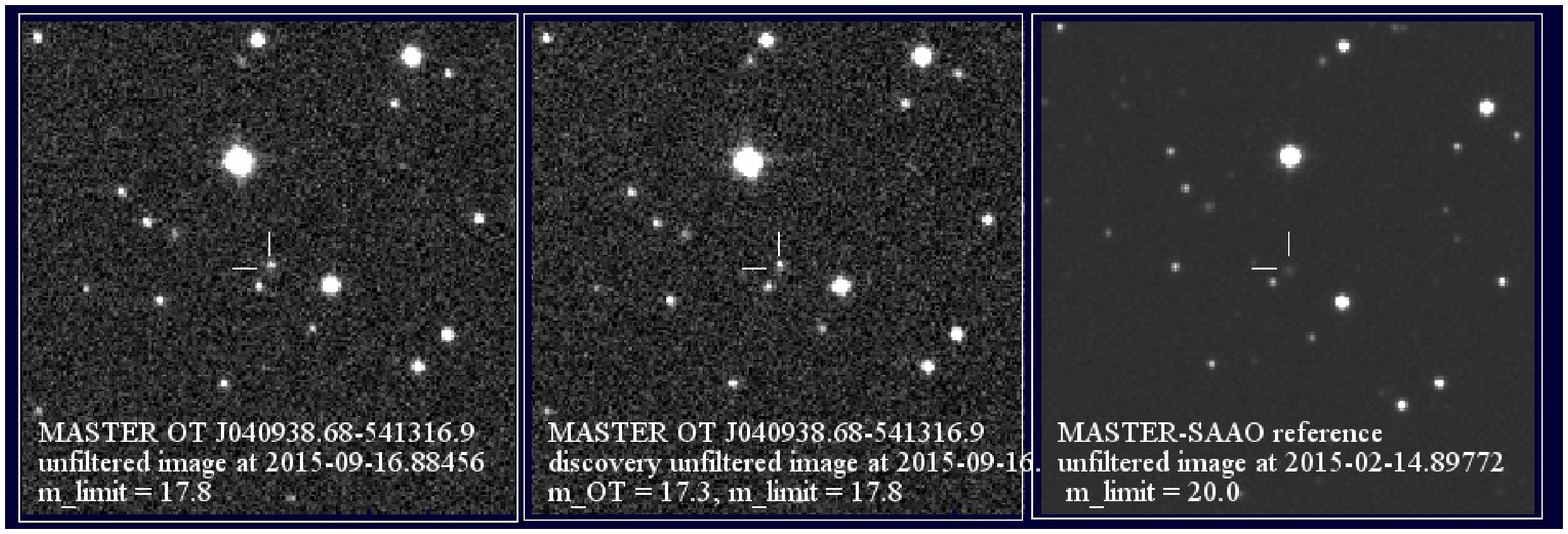}}
\caption{The discovery (left and middle) and reference (right) images for the supernova, MASTER OT J040938.68-541316.9, and associated with the z = 0.054 galaxy PGC421615, discovered by MASTER-SAAO inside the LIGO GW150914 error box during the first night of GW150914 inspection.}
\label{MASTER OTJ070747.72}

\end{figure*}

This OT was classified as a PSN, being located 0.9"W and 3.6"N of the center of the galaxy PGC421615 (Btot=18.4m). The discovery and reference images are shown in Figure \ref{MASTER OTJ070747.72}.

Later (after it was discovered on September 16, 2015) this supernova was observed several times in white (unfiltered) light during the regular survey performed by MASTER-SAAO telescope: on September 24, 2015; October 17, 18, and 19, 2015; November 25, 2015, and on January 26 and February 18, 2016. Several (three or more) frames with this sky area were available for each night, however, we analyzed only the combined images in order to increase the signal-to-noise ratio. We list the results of photometry in Table  \ref{tab:lc}. Formally, the supernova reached maximum light on September 24, 2015, however, the measured magnitude differed only slightly from the magnitude at the time of the discovery, and the error bars overlap. The supernova appears to have reached its maximum light between the observations of September 16 and October 24, 2015.

In order to study this supernova and its host galaxy in more detail, we took deep photometric images in \textit{B, g', r', i'} and \textit{z'} images of the area on 3 March 2016 with the SALTICAM CCD camera of the 10.4-meter Southern African Large Telescope (SALT) at the SAAO, as part of a spectroscopic followup program of MASTER OTs \citep{Buckley2016}. In some of these images the supernova can be seen clearly in different filters, 170 days after its discovery. We present the results of our photometric measurements in Table \ref{tab:lc}.

\begin{figure*}
\centering
\center{\includegraphics[width=1\linewidth]{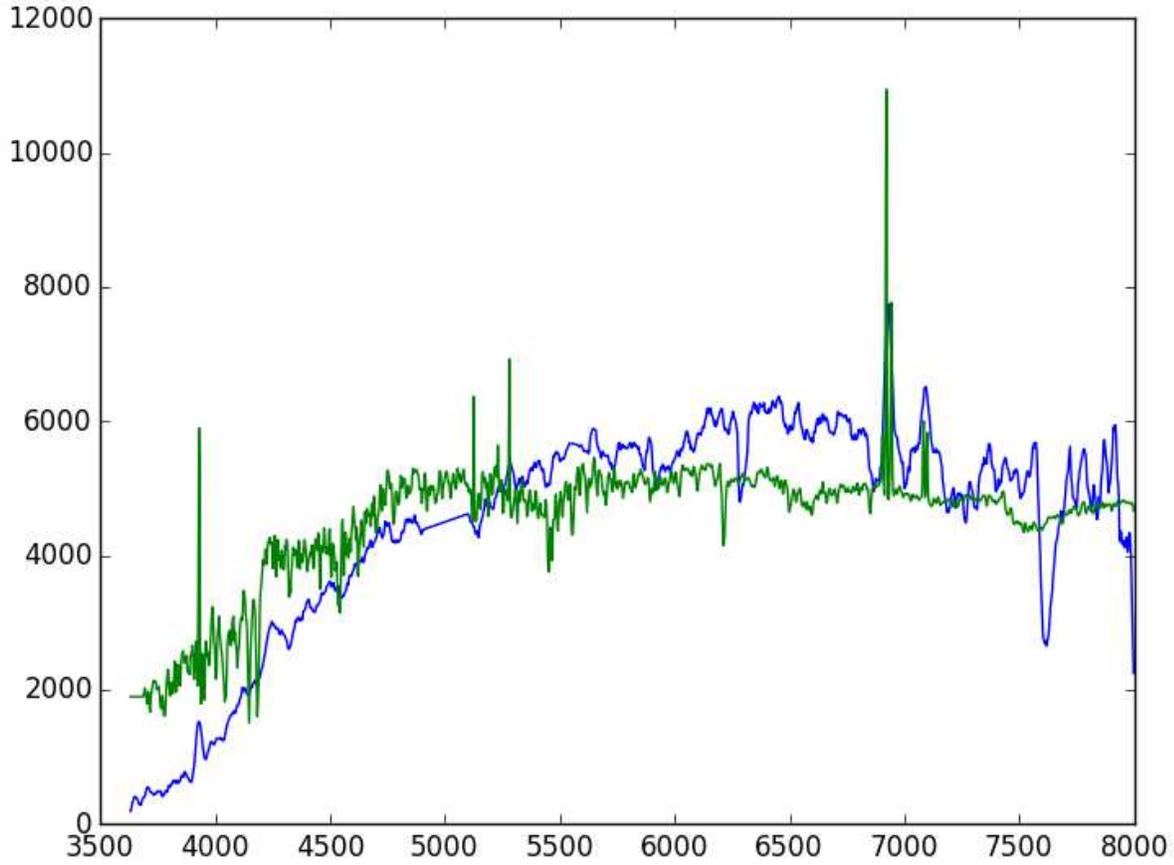}}
\caption{Figure showing the SALT spectrum of MASTER OT J040938.68-541316.9}
\label{spec}

\end{figure*}

Unfortunately, the scarcity of available photometric data prevents a determination of the supernova type. Photometric data listed in Table \ref{tab:lc} are consistent with the assumption that we are dealing with a type 1b/c or IIp supernova discovered near maximum light. The September 16 and 24, 2015 observations were evidently made near the maximum when the flux does not vary appreciably. Thus, given that supernovae generally reach their maximum light no later than 10 days after the explosion \footnote{https://c3.lbl.gov/nugent/nugent\_templates.html} and that the supernova considered was discovered near maximum light ~2d after GW150914, we can conclude that any association between GW150914 and our supernova is extremely unlikely. On 10 March 2016, we obtained low resolution ($\sim$300) spectrum, covering 340-1000nm, of the host galaxy PGC421615 in a 1800 s exposure.  The spectrum, shown in Figure \ref{spec}, has identified emission lines of [OII 3727], H$\alpha$ and [S II], resulting in a redshift determination of z=0.054, implying a relatively close galaxy.

\subsection{MASTER OT J070747.72-672205.6: A U Gem type Dwarf Nova}

The MASTER-SAAO auto-detection system discovered an OT source at (RA, Dec) = 07h 07m 47.72s -67d 22m 05.6s on 2015-09-21.99535 UT  ~\citep{Gress2015b}. The OT unfiltered magnitude was 16.9m (the limiting magnitude is 19.2m). The OT was seen in 8 images. There is no minor planet at this place. We have reference images without the OT taken on 2014-12-25.02683 UT and 2015-02-24.863UT with unfiltered magnitude limit of 20.0m and 20.3m, respectively.

\begin{figure*}
\center{\includegraphics[width=1\textwidth]{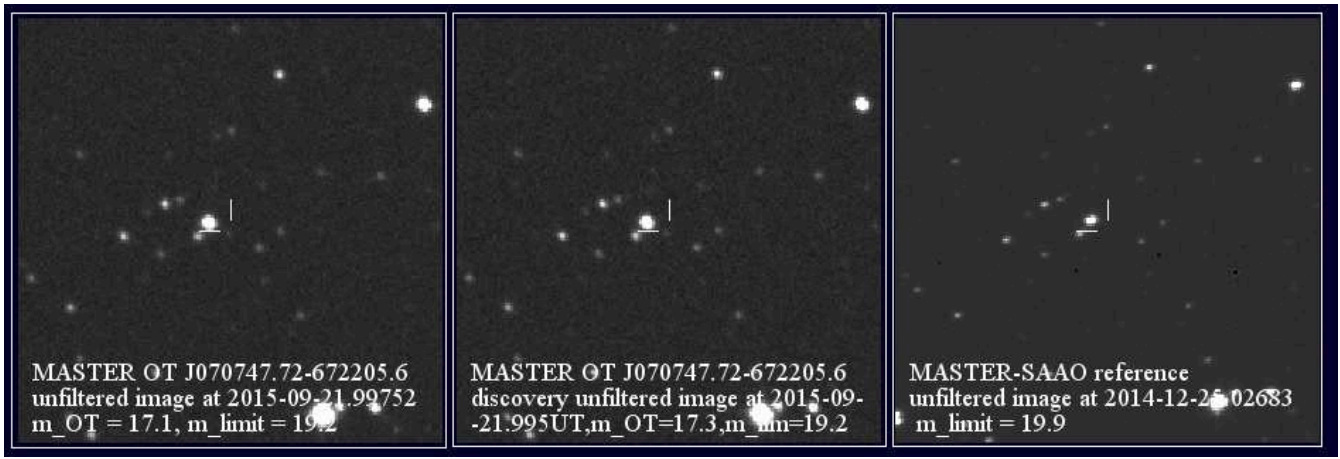}}
\caption{A U Gem type dwarf nova discovered by MASTER-SAAO inside the LIGO GW150914 error box. The discovery (left and middle) and reference (right) images of this OT are shown.}
\label{MASTER_OT_UGEM}

\end{figure*}

There is a USNO-B1 star (0226-0200013) 3.8" from the object with the blue and red magnitudes of B2=20.97 and R2=20.01, respectively, but AAVSO identified our OT with this star, namely a Cataclysmic Variable of the U Gem dwarf nova subclass (accreting white dwarf in binary system). The discovery and reference images are available at Figure \ref{MASTER_OT_UGEM}.

\subsection{MASTER OT J042822.91-604158.3 discovery: a Dwarf Nova}

The MASTER-SAAO auto-detection system discovered an OT source at (RA, Dec) = 04h 28m 22.91s -60d 41m 58.3s on 2015-09-16.90907 UT. The OT unfiltered magnitude is 18.2m (the limit is 19.2m). This OT was seen in 3 images on 2015-09-16 21:49:04.329 / 21:55:28.386  / 22:01:50.134UT, and is absent in the images on 2015-09-24 02:33:07 with $m_{lim}=19.6$. This implies the OT is not a supernova, despite being located close (18.7 arcsec)  to a galaxy (GALEXASC J042825.42-604155.3) with unknown redshift .

\begin{figure*}
\center{\includegraphics[width=1\textwidth]{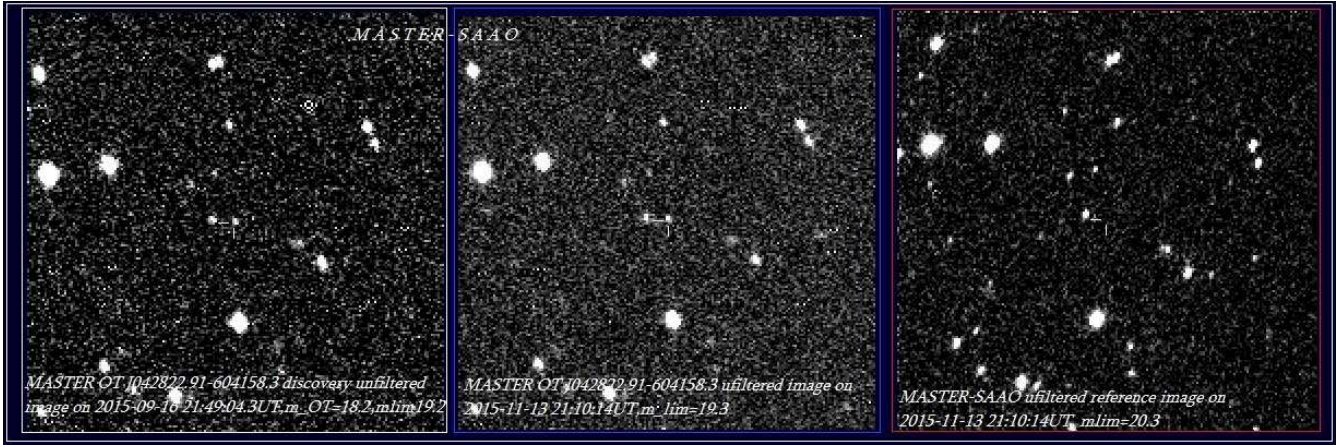}}
\caption{The discovery (left and middle) and reference (right) images for MASTER OT J042822.91-604158.3, a likely dwarf nova,  discovered by MASTER-SAAO inside the LIGO GW150914 error box during the first night of GW150914 inspection. }
\label{MASTER_OTJ042822.91}

\end{figure*}

\begin{figure*}
\centering
\center{\includegraphics[width=1\textwidth]{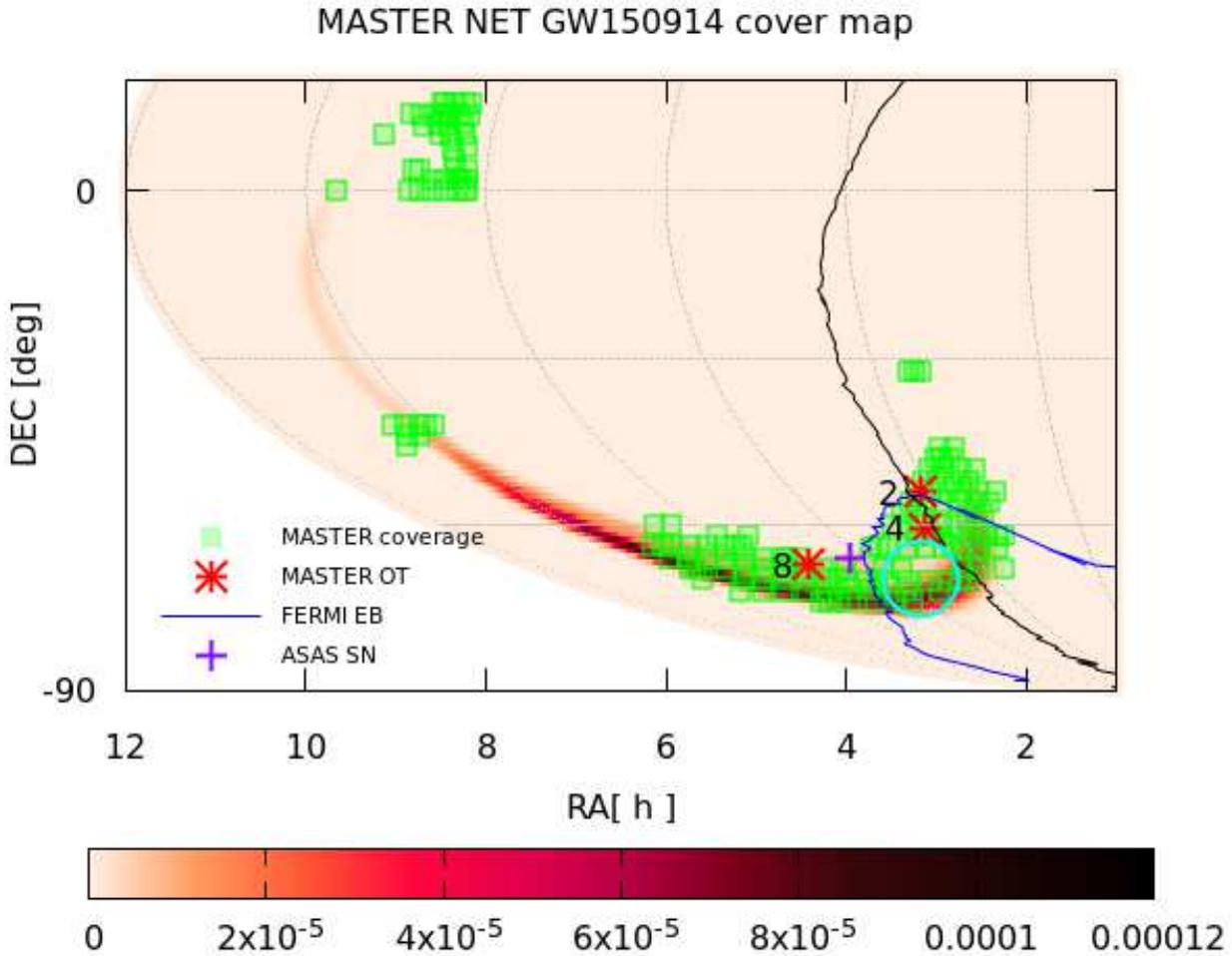}}
\caption{Positions of MASTER OTs compared with the Fermi error box (the blue line), Fermi Earth shadow (the black line), and the LIGO error region (the color palette). The color palette represents the probability distribution in the initial LIGO error region.  The green boxes show the distribution of the areas covered by MASTER net during the inspection of GW150914. This diagram shows only the fields with the probability greater than $1e^{-8}$.  The red asterisks indicate the optical transients discovered by MASTER during the LIGO error box inspections. These transients are described in detail in the paper and marked with bold font in Table \ref{tab:ligo1trans}. The light blue (cyan) circle shows the LMC region. The violet cross is  ASAS SN (\citep{ATEL8066}): another one  theoretically possible source of gravitational waves in  GW150914  error box, which, however, also is likely not associated with GW150914 event.}
\label{MASTER_FERMI}
\end{figure*}

We have reference images without the OT also taken on 2015-08-01 01:13:02UT with unfiltered magnitude limit of $m_{lim}=18.4$, and 2015-11-13 21:10:14UT with unfiltered $m_{lim}=20.3$, and on 2016-03-01 18:38:04UT with unfiltered $m_{lim}=21.3$.
There are no known sources in VIZIER database (22$^m$ POSS limit), so we suggest the amplitude of outburst is more then 3.8m. The most likely classification of this OT is a dwarf nova outburst, or possibly a dMe flare star. CRTS and NSVS also observed this area, but have never discovered any transient at this location. There are no minor planets at this location.
The discovery and reference images are shown in Figure \ref{MASTER_OTJ042822.91}.

\section{Fermi gamma-ray event}
\label{sec:fermi}
The participants of the program to search for electromagnetic counterparts of LIGO gravitational wave events included all X-ray and $\gamma$-ray observatories, including the Konus-Wind Russian-American experiment, the INTEGRAL, Swift and Fermi satellites and the MAXI experiment ~\citep{Abbott2016a}. However, Fermi was the only team to report the discovery of a very weak short (less than one second) $\gamma$-ray burst by GBM detector 0.4 seconds later the GW trigger ~\citep{Connaughton}. The burst had an energy of $\sim 3 \cdot 10^{-7} erg$ and was discovered \textit{post facto} in the archive record of the $\gamma$-ray background after receiving the G184098 alert. The last figure \ref{MASTER_FERMI} shows the localization domain of the Fermi event. Observations of the MASTER-SAAO telescope cover 90\% of the total area of the intersection of the LIGO and Fermi error regions. This area is covered only by MASTER observations and we found no traces of optical transients brighter than 19m that could be associated with the GW150914/G184098 event \citep{GCN18903}.

Let us now discuss the general possible connection between a $\gamma$-ray burst and a binary black-hole merger. We already pointed out that the emission of standard gamma-ray bursts is highly anisotropic and the probability of simultaneously recording gravitational waves and gamma-ray bursts is much less than 1/100. Furthermore, the luminosity of the gamma-ray burst, if we assume that it occurred at the same distance as the GW150914 event, can be estimated as $E_{Fermi} \sim 2 \cdot 10^{49}  erg/s$, which is much lower than the typical isotropic luminosity of $\gamma$-ray bursts. This hypothesis, which was actively discussed by Loeb (2016), has to be rejected.

Within the framework of standard general relativity, electromagnetic radiation in the case of a merger of two uncharged black holes could arise only because of the presence of extra matter in the binary black hole or in its immediate vicinity. For example, Lipunov and Sazhin noticed as far back as 1984 that a powerful electromagnetic burst could arise in the process of a merger of two supermassive black holes surrounded by a dense star cluster, which occur almost in all galactic nuclei. This is evidently impossible in the case of GW150914/G184098. However, a certain amount of mass could have accumulated around the black holes via accretion of interstellar gas during the pre-merger stage. This mass should be of about $\Delta M \sim 10^{-3} M_{\bigodot} $ if we adopt the typical energy relase factor of 10\% near accreting black holes. It is the typical mass of a Jupiter-like planet ~\citep{Cherep2016}. Although this may seem to be very small, given the $\Delta t \sim 0.4$ sec time lag corresponding to a distance of $c \Delta t \sim 10^{10} cm$, the plasma density near the black holes implied by this mass should be on the order of $\rho \sim \Delta M / (c \Delta t)^{3}  \sim 1 g / cm^{3}$, which is the density of Jupiter. However, such a ring of material or a planet is impossible to  imagine in a system of two blue supergiants (the progenitor to the black holes). A certain amount of matter could have been captured at the stage when the typical distance between the black holes was much smaller than $c \Delta t \sim 10^{10} cm$. Because of the continuous emission of gravitational waves the duration of this stage can not exceed


$$
t  \sim ( \frac{  I \Omega^{ 2}}{ 2L }) \sim 1 \;yr ( \frac {A}{10^{10}\;cm})^{4} / ( \frac {M}{60 M_{\bigodot}})^{3}
$$

Thus $t$ is about one year. The maximum mass that could have been accumulated over this year is $\Delta M \sim \dot{M} \cdot 1yr$, where the accretion rate can be estimated by the Bondi-Hoyle formula, see \citet{Lipunov1992}:


\begin{eqnarray*}
& \dot{M}  \approx \pi \frac{(2 G M)^{2}}{v^{3}} \rho  \sim \\
& \quad \sim 10^{-12} \; \frac{ M_{\bigodot}}{yr} \; (\frac{M}{60M_{\bigodot}})^{2} \; (\frac{\rho}{10^{-24} \;g/cm^{3}}) (\frac{V}{10 \; km/s})^{-3}
\end{eqnarray*}

where $M$ is the total mass of the black holes, $V$ is the velocity of the motion of black holes relative to the interstellar medium in the host galaxy, and $\rho$ is the density of the interstellar medium.

Obviously, the mass of $10^{-3} M_{\bigodot}$ cannot be accumulated in one year. So we conclud that the event discovered by Fermi $\gamma$ ray observatory is unrelated to the LIGO GW150914 event.

\section{Conclusion}

So  MASTER-Net of  telescopes robots was carried out the most extensive of all projects globally survey of  the first gravitational wave LIGO GW150914 event (see table 2 at \citep {Abbott2016b}). Despite the difficult conditions of observation (the error box was available for observation a few hours before sunrise), MASTER had covered 705 square degrees inside the initial (the LIB) and 590 square degrees inside the final (LALInf) error boxes. It should be noted that since the probability is nowhere equal to zero the area  outside 3-sigma  error box also can be taken into account. The observations outside 3-sigma  error box   were carried  out  while  the 3-sigma error box was unavailable for observation. In total, during the week after reports of GW150914 covered more than 5000 square degrees.

During the inspection of LIGO GW150914 event, MASTER-NET was found 8 optical transients (see table \ref {tab:ligo1trans}) 3 of which are located inside the 3-sigma of the initial and final square error. Of particular note is a supernova MASTER OT J070747.72s-541316.9, since it could theoretically give bursts of the gravitational wave.  However, the analysis (performed in the section~\ref{sec:sn}) indicates that the flash of this SN likely occurred earlier than LIGO GW150914 event and can not be associated with GW150914.  The others OTs also  can not be a called a LIGO GW150914 optical counterpart.


The common part of  LIGO and FERMI errors boxes, with deduction of the shadow of the Earth, is only a small area of ​​about 100 sq. degree in the southern hemisphere  (see fig. \ref {MASTER_FERMI}) which was almost completely ( $\sim 90 \%$)  covered  by the MASTER system. In this case, we also found no traces of optical transients brighter than $ 19-20^m $ that could be associated with the GW150914 / G184098 event.

\subsection*{Acknowledgments}

MASTER project is supported in part by the Development Programm of Lomonosov Moscow State University, Moscow Union OPTICA,  Russian Science Foundation 16-12-00085, Russian Foundation of Fundamental Research 15-02-07875, and National Research Foundation of South Africa.

We are especially  grateful to S.M.Bodrov for his long years MASTER's support.

We are grateful to Valerie Connaughton for the information about the FERMI GBM event.

\bibliographystyle{mn2e}
\bibliography{LigoMASTER}

\begin{thebibliography}{38}
\expandafter\ifx\csname natexlab\endcsname\relax\def\natexlab#1{#1}\fi

\bibitem[{{Abbott} {et~al}\mbox{.}(2016{\natexlab{a}}){Abbott}, {Abbott},
  {Abbott}, {Abernathy}, {Acernese}, {Ackley}, {Adams}, {Adams}, {Addesso},
  {Adhikari}, \& et~al.}]{Abbott2016c}
{Abbott} B.~P. {et~al.}, 2016{\natexlab{a}}, \apjl, 818, L22

\bibitem[{{Abbott} {et~al}\mbox{.}(2016{\natexlab{b}}){Abbott}, {Abbott},
  {Abbott}, {Abernathy}, {Acernese}, {Ackley}, {Adams}, {Adams}, {Addesso},
  {Adhikari}, \& et~al.}]{Abbott2016b}
---, 2016{\natexlab{b}}, ArXiv e-prints

\bibitem[{{Abbott} {et~al}\mbox{.}(2016{\natexlab{c}}){Abbott}, {Abbott},
  {Abbott}, {Abernathy}, {Acernese}, {Ackley}, {Adams}, {Adams}, {Addesso},
  {Adhikari}, \& et~al.}]{Abbott2016a}
---, 2016{\natexlab{c}}, Physical Review Letters, 116, 061102

\bibitem[{{Berger}, {Fong} \& {Chornock}(2013){Berger}, {Fong}, \&
  {Chornock}}]{Berger}
{Berger} E., {Fong} W., {Chornock} R., 2013, \apjl, 774, L23

\bibitem[{{Blinnikov} {et~al}\mbox{.}(1984){Blinnikov}, {Novikov},
  {Perevodchikova}, \& {Polnarev}}]{Blinnikov}
{Blinnikov} S.~I., {Novikov} I.~D., {Perevodchikova} T.~V., {Polnarev} A.~G.,
  1984, Pisma v Astronomicheskii Zhurnal, 10, 422

\bibitem[{{Buckley} {et~al}\mbox{.}(2015){Buckley}, {Breytenbach}, {Kniazev},
  {Kotze}, {Potter}, {Lipunov}, \& {Gorbovskoy}}]{Buckley2016}
{Buckley} D., {Breytenbach} H., {Kniazev} A., {Kotze} M.~M., {Potter} S.~B.,
  {Lipunov} V., {Gorbovskoy} E., 2015, in 3rd Annual Conference on High Energy
  Astrophysics in Southern Africa (HEASA2015), p.~3

\bibitem[{{Cherepashchuk}(2016)}]{Cherep2016}
{Cherepashchuk} A.~M., 2016, private communication

\bibitem[{{Connaughton} {et~al}\mbox{.}(2016){Connaughton}, {Burns},
  {Goldstein}, {Briggs}, {Zhang}, {Hui}, {Jenke}, {Racusin}, {Wilson-Hodge},
  {Bhat}, {Bissaldi}, {Cleveland}, {Fitzpatrick}, {Giles}, {Gibby}, {Greiner},
  {von Kienlin}, {Kippen}, {McBreen}, {Mailyan}, {Meegan}, {Paciesas},
  {Preece}, {Roberts}, {Sparke}, {Stanbro}, {Toelge}, {Veres}, {Yu}, \&
  {authors}}]{Connaughton}
{Connaughton} V. {et~al.}, 2016, ArXiv e-prints

\bibitem[{{Eichler} {et~al}\mbox{.}(1989){Eichler}, {Livio}, {Piran}, \&
  {Schramm}}]{Eichler}
{Eichler} D., {Livio} M., {Piran} T., {Schramm} D.~N., 1989, \nat, 340, 126

\bibitem[{{Gorbovskoy} {et~al}\mbox{.}(2016){Gorbovskoy}, {Lipunov}, {Buckley},
  {Kornilov}, {Balanutsa}, {Tyurina}, {Kuznetsov}, {Kuvshinov}, {Gorbunov},
  {Vlasenko}, {Popova}, {Chazov}, {Potter}, {Kotze}, {Kniazev}, {Gress},
  {Budnev}, {Ivanov}, {Yazev}, {Tlatov}, {Senik}, {Dormidontov}, {Parhomenko},
  {Krushinski}, {Zalozhnich}, {Castro-Tirado}, {S{\'a}nchez-Ram{\'{\i}}rez},
  {Sergienko}, {Gabovich}, {Yurkov}, {Levato}, {Saffe}, {Mallamaci}, {Lopez},
  {Podest}, \& {Vladimirov}}]{Gorbovskoy2016}
{Gorbovskoy} E.~S. {et~al.}, 2016, \mnras, 455, 3312

\bibitem[{{Gorbovskoy} {et~al}\mbox{.}(2013){Gorbovskoy}, {Lipunov},
  {Kornilov}, {Belinski}, {Kuvshinov}, {Tyurina}, {Sankovich}, {Krylov},
  {Shatskiy}, {Balanutsa}, {Chazov}, {Kuznetsov}, {Zimnukhov}, {Shumkov},
  {Shurpakov}, {Senik}, {Gareeva}, {Pruzhinskaya}, {Tlatov}, {Parkhomenko},
  {Dormidontov}, {Krushinsky}, {Punanova}, {Zalozhnyh}, {Popov}, {Burdanov},
  {Yazev}, {Budnev}, {Ivanov}, {Konstantinov}, {Gress}, {Chuvalaev}, {Yurkov},
  {Sergienko}, {Kudelina}, {Sinyakov}, {Karachentsev}, {Moiseev}, \&
  {Fatkhullin}}]{Gorbovskoy2013}
---, 2013, Astronomy Reports, 57, 233

\bibitem[{{Gorbovskoy} {et~al}\mbox{.}(2012){Gorbovskoy}, {Lipunova},
  {Lipunov}, {Kornilov}, {Belinski}, {Shatskiy}, {Tyurina}, {Kuvshinov},
  {Balanutsa}, {Chazov}, {Kuznetsov}, {Zimnukhov}, {Kornilov}, {Sankovich},
  {Krylov}, {Ivanov}, {Chvalaev}, {Poleschuk}, {Konstantinov}, {Gress},
  {Yazev}, {Budnev}, {Krushinski}, {Zalozhnich}, {Popov}, {Tlatov},
  {Parhomenko}, {Dormidontov}, {Senik}, {Yurkov}, {Sergienko}, {Varda},
  {Kudelina}, {Castro-Tirado}, {Gorosabel}, {S{\'a}nchez-Ram{\'{\i}}rez},
  {Jelinek}, \& {Tello}}]{Gorbovskoy2012}
---, 2012, \mnras, 421, 1874

\bibitem[{{Gress} {et~al}\mbox{.}(2015{\natexlab{a}}){Gress}, {Balanutsa},
  {Lipunov}, {Gorbovskoy}, {Buckley}, {Rebolo}, {Tyurina}, {Kornilov},
  {Kuznetsov}, {Gorbunov}, {Vlasenko}, {Chazov}, {Potter}, {Kotze},
  {Serra-Ricart}, {Lodieu}, {Israelian}, {Budnev}, {Ivanov}, {Tlatov},
  {Dormidontov}, {Senik}, {Sergienko}, {Gabovich}, {Yurkov}, {Krushinski}, \&
  {Zalozhnykh}}]{Gress2015a}
{Gress} O. {et~al.}, 2015{\natexlab{a}}, The Astronomer's Telegram, 8065

\bibitem[{{Gress} {et~al}\mbox{.}(2015{\natexlab{b}}){Gress}, {Balanutsa},
  {Lipunov}, {Gorbovskoy}, {Samus}, {Kornilov}, {Tiurina}, {Kuznetsov},
  {Gorbunov}, {Vlasenko}, {Potter}, {Kotze}, {Budnev}, {Ivanov}, {Tlatov},
  {Dormidontov}, {Sergienko}, {Gabovich}, {Yurkov}, {Krushinskiy}, {Rebolo},
  {Serra-Ricart}, {Lodieu}, \& {Israelian}}]{Gress2015b}
---, 2015{\natexlab{b}}, The Astronomer's Telegram, 8087

\bibitem[{{Hansen} \& {Lyutikov}(2001)}]{Hansen}
{Hansen} B.~M.~S., {Lyutikov} M., 2001, \mnras, 322, 695

\bibitem[{{Kornilov} {et~al}\mbox{.}(2012){Kornilov}, {Lipunov}, {Gorbovskoy},
  {Belinski}, {Kuvshinov}, {Tyurina}, {Shatsky}, {Sankovich}, {Krylov},
  {Balanutsa}, {Chazov}, {Kuznetsov}, {Zimnuhov}, {Senik}, {Tlatov},
  {Parkhomenko}, {Dormidontov}, {Krushinsky}, {Zalozhnyh}, {Popov}, {Yazev},
  {Budnev}, {Ivanov}, {Konstantinov}, {Gress}, {Chvalaev}, {Yurkov},
  {Sergienko}, \& {Kudelina}}]{Kornilov2012}
{Kornilov} V.~G. {et~al.}, 2012, Experimental Astronomy, 33, 173

\bibitem[{{Li} \& {Paczy{\'n}ski}(1998)}]{Li}
{Li} L.-X., {Paczy{\'n}ski} B., 1998, \apjl, 507, L59

\bibitem[{{Lipunov} {et~al}\mbox{.}(2016{\natexlab{a}}){Lipunov}, {Gorbovskoy},
  {Afanasiev}, {Tatarnikova}, {Denisenko}, {Makarov}, {Tiurina}, {Krushinsky},
  {Vinokurov}, {Balanutsa}, {Kuznetsov}, {Gress}, {Sergienko}, {Yurkov},
  {Gabovich}, {Tlatov}, {Senik}, {Vladimirov}, \& {Popova}}]{Lipunov2016AA}
{Lipunov} V. {et~al.}, 2016{\natexlab{a}}, \aap, 588, A90

\bibitem[{{Lipunov}, {Gorbovskoy} \& {Buckley}(2015){Lipunov}, {Gorbovskoy}, \&
  {Buckley}}]{GCN18390}
{Lipunov} V., {Gorbovskoy} E., {Buckley} D., 2015, GRB Coordinates Network,
  18390

\bibitem[{{Lipunov} {et~al}\mbox{.}(2015){Lipunov}, {Gorbovskoy}, {Tyurina},
  {Kornilov}, {Balanutsa}, {Kuznetsov}, {Kuvshinov}, {Buckley}, {Potter},
  {Kniazev}, {Kotze}, {Ivanov}, {Yazev}, {Budnev}, {Gres}, {Chuvalaev},
  {Poleshchuk}, {Tlatov}, {Parhomenko}, {Dormidontov}, {Sennik}, {Yurkov},
  {Sergienko}, {Varda}, {Krushinski}, {Zalozhnih}, {Popov}, {Levato}, {Saffe},
  {Mallamaci}, {Lopez}, \& {Podest}}]{GCN18333}
{Lipunov} V. {et~al.}, 2015, GRB Coordinates Network, 18333

\bibitem[{{Lipunov} {et~al}\mbox{.}(2016{\natexlab{b}}){Lipunov}, {Gorbovskoy},
  {Tyurina}, {Kornilov}, {Balanutsa}, {Kuznetsov}, {Kuvshinov}, {Buckley},
  {Potter}, {Kniazev}, {Kotze}, {Ivanov}, {Yazev}, {Budnev}, {Gres},
  {Chuvalaev}, {Poleshchuk}, {Tlatov}, {Parhomenko}, {Dormidontov}, {Sennik},
  {Yurkov}, {Sergienko}, {Varda}, {Krushinski}, {Zalozhnih}, {Popov}, {Levato},
  {Saffe}, {Mallamaci}, {Lopez}, \& {Podest}}]{GCN18903}
---, 2016{\natexlab{b}}, GRB Coordinates Network, 18903

\bibitem[{{Lipunov} {et~al}\mbox{.}(2010){Lipunov}, {Kornilov}, {Gorbovskoy},
  {Shatskij}, {Kuvshinov}, {Tyurina}, {Belinski}, {Krylov}, {Balanutsa},
  {Chazov}, {Kuznetsov}, {Kortunov}, {Sankovich}, {Tlatov}, {Parkhomenko},
  {Krushinsky}, {Zalozhnyh}, {Popov}, {Kopytova}, {Ivanov}, {Yazev}, \&
  {Yurkov}}]{Lipunov2010}
---, 2010, Advances in Astronomy, 2010, 349171

\bibitem[{{Lipunov} {et~al}\mbox{.}(2016{\natexlab{c}}){Lipunov}, {Kornilov},
  {Gorbovskoy}, {Tyurina}, {Balanutsa}, {Kuznetsov}, {Kuvshinov}, {Buckley},
  {Potter}, {Kniazev}, {Kotze}, {Ivanov}, {Yazev}, {Budnev}, {Gres},
  {Chuvalaev}, {Poleshchuk}, {Tlatov}, {Parhomenko}, {Dormidontov}, {Sennik},
  {Yurkov}, {Sergienko}, {Varda}, {Krushinski}, {Zalozhnih}, {Popov}, {Levato},
  {Saffe}, {Mallamaci}, {Lopez}, \& {Podest}}]{MASTER_LIGO_Theory}
---, 2016{\natexlab{c}}, in MNRAS, in print

\bibitem[{{Lipunov}(1992)}]{Lipunov1992}
{Lipunov} V.~M., 1992, Springer, NewYork/Berlin

\bibitem[{{Lipunov}, {Bogomazov} \& {Abubekerov}(2005){Lipunov}, {Bogomazov},
  \& {Abubekerov}}]{Lipunov2005}
{Lipunov} V.~M., {Bogomazov} A.~I., {Abubekerov} M.~K., 2005, \mnras, 359, 1517

\bibitem[{{Lipunov} \& {Gorbovskoy}(2008)}]{Lipunov2008}
{Lipunov} V.~M., {Gorbovskoy} E.~S., 2008, \mnras, 383, 1397

\bibitem[{{Lipunov} {et~al}\mbox{.}(2016{\natexlab{d}}){Lipunov}, {Gorosabel},
  {Pruzhinskaya}, {Postigo}, {Pelassa}, {Tsvetkova}, {Sokolov}, {Kann}, {Xu},
  {Gorbovskoy}, {Krushinski}, {Kornilov}, {Balanutsa}, {Boronina}, {Budnev},
  {Cano}, {Castro-Tirado}, {Chazov}, {Connaughton}, {Delvaux}, {Frederiks},
  {Fynbo}, {Gabovich}, {Goldstein}, {Greiner}, {Gress}, {Ivanov}, {Jakobsson},
  {Klose}, {Knust}, {Komarova}, {Konstantinov}, {Krylov}, {Kuvshinov},
  {Kuznetsov}, {Lipunova}, {Moskvitin}, {Pal'shin}, {Pandey}, {Poleshchuk},
  {Schmidl}, {Sergienko}, {Sinyakov}, {Schulze}, {Sokolov}, {Sokolova},
  {Sparre}, {Th{\"o}ne}, {Tlatov}, {Tyurina}, {Ulanov}, {Yazev}, \&
  {Yurkov}}]{Lipunov2016}
{Lipunov} V.~M. {et~al.}, 2016{\natexlab{d}}, \mnras, 455, 712

\bibitem[{{Lipunov} {et~al}\mbox{.}(2004){Lipunov}, {Krylov}, {Kornilov},
  {Borisov}, {Kuvshinov}, {Belinsky}, {Kuznetsov}, {Potanin}, {Antipov},
  {Tyurina}, {Gorbovskoy}, \& {Chilingaryan}}]{Lipunov2004}
---, 2004, Astronomische Nachrichten, 325, 580

\bibitem[{{Lipunov} \& {Panchenko}(1996)}]{Lipunov1996a}
{Lipunov} V.~M., {Panchenko} I.~E., 1996, \aap, 312, 937

\bibitem[{{Lipunov}, {Postnov} \& {Prokhorov}(1997{\natexlab{a}}){Lipunov},
  {Postnov}, \& {Prokhorov}}]{Lipunov1997a}
{Lipunov} V.~M., {Postnov} K.~A., {Prokhorov} M.~E., 1997{\natexlab{a}},
  Astronomy Letters, 23, 492

\bibitem[{{Lipunov}, {Postnov} \& {Prokhorov}(1997{\natexlab{b}}){Lipunov},
  {Postnov}, \& {Prokhorov}}]{Lipunov1997b}
---, 1997{\natexlab{b}}, \na, 2, 43

\bibitem[{{Lipunov}, {Postnov} \& {Prokhorov}(1997{\natexlab{c}}){Lipunov},
  {Postnov}, \& {Prokhorov}}]{Lipunov1997c}
---, 1997{\natexlab{c}}, \mnras, 288, 245

\bibitem[{{Lipunova} \& {Lipunov}(1998)}]{Lipunova}
{Lipunova} G.~V., {Lipunov} V.~M., 1998, \aap, 329, L29

\bibitem[{{Metzger} {et~al}\mbox{.}(2010){Metzger}, {Mart{\'{\i}}nez-Pinedo},
  {Darbha}, {Quataert}, {Arcones}, {Kasen}, {Thomas}, {Nugent}, {Panov}, \&
  {Zinner}}]{Metzger}
{Metzger} B.~D. {et~al.}, 2010, \mnras, 406, 2650

\bibitem[{{Narayan}, {Paczynski} \& {Piran}(1992){Narayan}, {Paczynski}, \&
  {Piran}}]{Narayan}
{Narayan} R., {Paczynski} B., {Piran} T., 1992, \apjl, 395, L83

\bibitem[{{Nicholls} {et~al}\mbox{.}(2015){Nicholls}, {Brown}, {Stanek},
  {Holoien}, {Kochanek}, {Rivera}, {Simonian}, {Basu}, {Beacom}, {Thompson},
  {Shappee}, {Prieto}, {Bersier}, {Dong}, {Chen}, {Brimacombe}, {Falco},
  {Wozniak}, {Pojmanski}, {Bock}, {Conseil}, {Fernandez}, {Kiyota}, {Marples},
  \& {Masi}}]{ATEL8066}
{Nicholls} B. {et~al.}, 2015, The Astronomer's Telegram, 8066

\bibitem[{{Singer}(2016)}]{GCN18330}
{Singer} L., 2016, GRB Coordinates Network, 18330

\bibitem[{{Tanvir} {et~al}\mbox{.}(2013){Tanvir}, {Levan}, {Fruchter},
  {Hjorth}, {Hounsell}, {Wiersema}, \& {Tunnicliffe}}]{Tanvir}
{Tanvir} N.~R., {Levan} A.~J., {Fruchter} A.~S., {Hjorth} J., {Hounsell} R.~A.,
  {Wiersema} K., {Tunnicliffe} R.~L., 2013, \nat, 500, 547

\end{thebibliography}
\newpage
\onecolumn

\begin{table}
\small
\caption{Coordinates of the observatories comprising the MASTER-NET. \label{tab:masternet}}
\bigskip
\centering
\begin{tabular}{lrrrcl}
\hline\hline
Site  & Longitude $\lambda$ & latitude $\phi$ & height $h$ & Country & Note \\
\hline
MASTER-Amur       & $+08^h29^m56^s\!.0$  & $+50^\circ 19^\prime 07^{\prime\prime}$ & 215 & Russia & Blagoveshchensk, Far East\\
MASTER-Tunka      & $+06^h52^m16^s\!.1$  & $+51^\circ 48^\prime 34^{\prime\prime}$ & 700 & Russia & Tunka valley  \\
MASTER-Ural   & $+03^h58^m11^s\!.2$  & $+57^\circ 02^\prime 13^{\prime\prime}$ & 290 & Russia & Ural, Kourovka  \\
MASTER-Kislovodsk & $+02^h50^m04^s\!.0$  & $+43^\circ 45^\prime 00^{\prime\prime}$ & 2067& Russia & Caucasus mountains\\
MASTER-SAAO & $+01^h23^m14^s\!.7$  & $-32^\circ 22^\prime 49^{\prime\prime}$ & 1760 &  South Africa & South African Astronomical Observatory \\
MASTER-IAC & $-01^h06^m02^s\!.5$  & $+28^\circ 17^\prime 55^{\prime\prime}$ & 2422 &  Spain & Canary Islands, Teide Observatory \\
MASTER-ICATE & $-04^h37^m18^s\!.3$  & $-31^\circ 48^\prime 08^{\prime\prime}$ & 2430 &  Argentina & wide field cameras only \\

\hline\hline
\end{tabular}
\end{table}

\begin{table*}
 \centering
  \caption{MASTER NET survey parameters during the GW150914 inspection. The listed parameters include the covered sky area, the covered sky area inside the GW150914 error box, and the total contained probability for all four possible GW150914 localizations. cWB, LIB, BSTR and  LALInf are abbreviations of different variants  of LIGO data processings, for event localization. They are described in detail in the paper \citep{Abbott2016b} paragraph 2, page 14 and references therein.\label {tab:MASTER_Prob} }
  \begin{tabular}{@{}lccccccccc@{}}
  \hline
  \hline
Site&Area Full &Area in final  EB&\multicolumn{4}{c}{Contained probability (\%)}\\
&(deg2)& (deg2)&cWB&LIB&BSTR.&LALInf  .\\
 \hline
 MASTER-NET	&5246&590&56&35&55&49\\
 MASTER-SAAO&1072&496&55&33&55&49\\
MASTER-Kislovodsk&1504&84&1.1&$\leq 1 e{-3}$&$\leq 1 e{-3}$&$\leq 1 e{-3}$\\
MASTER-Tunka&990&28&0.9&$\leq 1 e{-3}$&$\leq 1 e{-3}$&$\leq 1 e{-3}$\\
MASTER-IAC&1587&24&1.0&$\leq 1 e{-3}$&$\leq 1 e{-3}$&$\leq 1 e{-3}$\\
MASTER-Amur&438&0&$\leq 1 e{-3}$&$\leq 1 e{-3}$&$\leq 1 e{-3}$&$\leq 1 e{-3}$\\
MASTER-Ural&261&0&$\leq 1 e{-3}$&$\leq 1 e{-3}$&$\leq 1 e{-3}$&$\leq 1 e{-3}$\\
\hline
\end{tabular}
\end{table*}

\begin{table*}
 \centering
  \caption{Optical transients  discovered by MASTER-Net autodetection system  during GW150914 observations. The discovery date (``Date/UT'' column) is expressed as the day number in September 2015. For example, 16.017 means that the transient was discovered at 2015-09-16.017 UT. Type is the transient type: SN - supernova,  PSN - probable supernova, DN - dwarf nova outburst. Mag is the unfiltered magnitude of the transient defineded as $0.8 \times R_2 + 0.2 \times B_2 $, where R and B are the corresponding USNO B1.0 catalog magnitudes. Site indicates the particular observatory of the MASTER net that discovered the transient. Ligo Prob is the Ligo probability at the OT location in the sky.\label {tab:ligo1trans}}
  \begin{tabular}{@{}lccccccccc@{}}
  \hline
  \hline
N&OT Name&Date,UT&Type&Mag&ATel&MASTER-Site&Ligo Prob.&Comment\\
  \hline
1&MASTER OT J010654.20-254135.1&16.017&DN&18.3&8087&SAAO&$\leq 10^{-42}$&$Ampl>2m$\\
2&MASTER OT J040938.68-541316.9&16.879&SN&17.3&8065&SAAO&$1.9 \cdot10^{-6}$&PSN in PGC421615\\
3&MASTER OT J183934.91+414404.2&16.890&SN&17.2&8064&IAC	&$1.3 \cdot10^{-37}$ &Ultraluminous PSN\\
4&MASTER OT J042822.91-604158.3&16.909&DN&18.2&-&SAAO&$3.6\cdot10^{-6}$&$Ampl>3.8m$\\
5&MASTER OT J202411.65-172512.5&17.862&DN&17.9&8065&IAC&$8.2\cdot10^{-42}$&$Ampl>2.7m$\\
6&MASTER OT J040140.85+670613.9&19.848&SN&17.7&8075&Tunka&$\leq 10^{-42}$&PSN in PGC2695052$^{*}$\\
7&MASTER OT J092544.53+341636.1&21.072&SN&15.6&8077&Kislovodsk&$9.3 \cdot 10^{-41}$ &SN2015aq(II,UGC05015)\\
8&MASTER OT J070747.72-672205.6&21.995&DN&16.9&8087&SAAO&$8.4 \cdot 10^{-6}$& $Ampl>3.4m$\\
  \hline
\end{tabular}
* PSN in PGC2695052 was discovered by MASTER auto-detection system during the Fermi trigger 464366002 inspection.

\end{table*}

\begin{table*}
 \centering
  \caption{The brightness measurements for the supernova MASTER OT J040938.68-541316.9. \label {tab:lc}}
  \begin{tabular}{@{}lccccccccc@{}}
 \hline
Observatory, Instrument & Band& JD& Exptime& Mag &Err. Mag \\
  \hline
MASTER-SAAO& C&2457282.384&480&17.2&0.1\\
MASTER-SAAO& C&2457289.597&180&17.1&0.1\\
MASTER-SAAO& C&2457312.469&1620&18.0&0.2\\
MASTER-SAAO& C&2457313.470&2160&18.0&0.2\\
MASTER-SAAO& C&2457314.503&720&18.1&0.2\\
MASTER-SAAO& C&2457352.453&180& $\le 19.2$\\	
MASTER-SAAO& C&2457414.438&180& $\le 20.2$\\	
MASTER-SAAO& C&2457437.308&13500& $\le 20.6$\\	
SALTICAM& r'&2457452.309&100&20.0&0.1\\
SALTICAM& g'&2457452.312&100&20.6&0.1\\
SALTICAM& B&2457452.313&100&21.2&0.1\\
  \hline
\end{tabular}
\end{table*}

%
%
 \end{document}